\documentstyle[apjfonts,emulateapj]{article}

\begin{document}
\submitted{Preprint: The Astrophysical Journal (1998), in press}
\title{The simultaneous spectrum of Sgr A* from $\lambda$20cm to
$\lambda$1mm and the nature of the mm-excess}
\author{Heino Falcke\altaffilmark{1,2}, W.M.~Goss\altaffilmark{3},
Hiroshi Matsuo\altaffilmark{4}, Peter~Teuben\altaffilmark{1}, 
Jun-Hui~Zhao\altaffilmark{5}, Robert~Zylka\altaffilmark{6}}

\altaffiltext{1}{Astronomy Department, University of Maryland, College Park, MD
20742-2421 (hfalcke,teuben@astro.umd.edu)}
\altaffiltext{2}{Current Address: Max-Planck-Institut f\"ur
Radioastronomie, Auf dem H\"ugel 69, D-53121 Bonn, Germany (hfalcke@mpifr-bonn.mpg.de)}
\altaffiltext{3}{NRAO, P.O. Box O, Socorro, NM 87801-0387 (mgoss@nrao.edu)}
\altaffiltext{4}{Nobeyama Radio Observatory, Minamimaki, Minamisaku, 
Nagano 384-13, Japan (matsuo@nro.nao.ac.jp)}
\altaffiltext{5}{Center for Astrophysics, 60 Garden Street, Cambridge,
MA 02138 (jzhao@cfa.harvard.edu)}
\altaffiltext{6}{Institut f\"ur Theoretische Astrophysik, Tiergartenstra\ss{}e
15, D-69121 Heidelberg, Germany  (rzylka@ita.uni-heidelberg.de)}

\begin{abstract}
We report results from a multiwavelength campaign to measure the
simultaneous spectrum of the super-massive black hole candidate Sgr A*
in the Galactic Center from cm to mm-wavelengths using the VLA, BIMA,
the Nobeyama 45m, and the IRAM 30m telescopes. The observations confirm
that the previously detected mm-excess is an intrinsic feature of the
spectrum of Sgr~A*. The excess can be interpreted as due to the
presence of an ultra-compact component of relativistic plasma with a
size of a few Schwarzschild radii near the black hole. If so, Sgr~A*
might offer the unique possibility to image the putative black hole
against the background of this component with future mm-VLBI
experiments.
\end{abstract}

\keywords{galaxies: nuclei---Galaxy: Center---black hole physics}

\section{Introduction}
It is now almost certain that the Galaxy hosts a dark mass of
$2.6\cdot10^6M_\odot$ in its very center. This has been demonstrated
most convincingly by high-resolution near-infrared speckle imaging of
the central star cluster (Eckart \& Genzel 1996) where proper motions
of stars yield velocities of up to $\sim2000$ km/sec near the compact
radio source Sgr A*. Its location within the optical frame has now
been determined with high accuracy by Menten et al.~(1996). To explain
the nature of Sgr A* various models have been proposed, such as an
advection dominated accretion flow (Narayan et al.~1995), Bondi-Hoyle
accretion (Melia 1994), and a jet outflow (Falcke et al.~1993; Falcke
1996a)---all involving a supermassive black hole.

And indeed, the gravitational potential required to explain the run of
stellar velocities around Sgr A* is that of a point source within
$\sim0.01$ pc, implying a central dark mass density of
$>10^{12}M_\odot {\rm pc}^{-3}$. Such a mass density is too high to be
explained by a cluster of stellar remnants and therefore strongly
supports the presence of a black hole (Maoz 1997) in the Milky
Way. Nevertheless, a definite and direct proof for the existence of
black holes in general and in the Galactic Center in particular is
still missing, because the scales probed by NIR imaging are still
large compared to the the event horizon, the defining characteristic
of a black hole.

However, since the black hole is most certainly associated with the
non-thermal radio source Sgr A* (e.g. Eckart et al. 1993, Backer
1996), interferometric observations at mm-wavelengths might eventually
have the resolution to reach even those extreme scales. Hence the
mm-spectrum of Sgr~A* is of considerable interest. Early
mm-observations of Sgr~A* suggested the presence of a mm-excess due to
a sub-mm bump (Zylka et al.~1992 \& 1995) in the spectrum, possibly
caused by an ultra-compact region in Sgr A* (Falcke et al.~1993;
Falcke 1996b). The existence of this bump was, however, uncertain due
to the variability of Sgr A*. In order to determine the reality of
this crucial feature in the spectrum, we have conducted
an experiment to measure the spectrum of Sgr~A* simultaneously from
$\lambda$20cm to $\lambda$1mm.

The Galactic Center (GC) was observed on three consecutive days on
25-27 October 1996 with four different telescopes (VLA, BIMA, Nobeyama
45 m, \& IRAM 30 m) on three continents. The campaign was set up with
redundancies in wavelengths and time coverage. The observations were
prepared, performed, and reduced independently by four different
groups and will be described in Section 2. The results are presented
in Sec.~3, and their implications are discussed in Sec.~4.  For a
discussion of the overall spectrum of Sgr~A* and a comparison to model
predictions, we refer the reader to the recent work by Serabyn et
al.~(1997). Here we will solely focus on the mm-excess and its
implications.

\section{Observations and data reduction}
\subsection{VLA A-configuration}
Observations of Sgr A* with the VLA in its A-configuration were
scheduled for 2 hrs on each of three successive days on 25-27 October
1996. Sgr A* was observed at 43.3 GHz for the entire run on each of
the dates using the thirteen antennas mounted with the $\lambda$7mm
receivers. The rest of the array was used to observe Sgr A* at 22.5,
14.9, 8.45, 4.85, 1.64, 1.44, and 1.36 GHz in snapshot mode
($\sim$10 min observing time on the target source at each
frequency). The weather was cloudy during the
observations.  On October 25th all data were lost due to strong
wind and on the 26th data at 5 and 1.67 GHz were lost due
to setup problems. 

The flux density scale was determined using the primary calibrator 3C
286 assuming flux densities of 1.45, 2.50, 3.43, 5.18, 7.48, 13.8,
14.7, and 15.0 Jy for the frequencies given above (from 43.3 to 1.36
GHz). Both 1730--130 and 1741--312 were used to check the phases and
amplitudes. At 43 GHz the AIPS task {\sc Elint} was used to correct for
elevation-dependent antenna gains. The data were further
self-calibrated for the phase only, images were reconstructed, and
flux densities were measured in the image domain. Uncertainties (1
$\sigma$) of the flux densities were estimated including the
r.m.s. and the uncertainties in the flux density calibration.


\subsection{BIMA C-configuration}
BIMA observed the GC in its C-configuration on all three days under
good weather conditions throughout. Uranus was initially used as the
primary flux calibrator. The observing sequence followed a triangular
pattern including Sgr A*, NRAO 530 (1730-130), and OH 5.89--0.39 (at
17 57 26.7 -24 03 54 (B1950), a luminous compact \ion{H}{2} region),
with 5 min, 2.5 min and 1.7 min integration times respectively.  The
total observing time each day was 1.5 - 2 hours.

Since the data was decorrelated at the long baselines, NRAO 530 was
not a suitable calibrator. Instead we used OH 5.89--0.39 as a phase
and amplitude calibrator for Sgr A* which is only $\sim7^\circ$
away. The flux density of OH5.89--0.39 was taken to be 8.5 Jy at
$\lambda$3mm as determined by the Nobeyama single dish.

Since OH 5.89--0.39 is slightly resolved at the longest baselines
(disk of $\sim$ 5\arcsec, e.g. Zijlstra et al. 1990) we only used
baselines shortwards of 10 K$\lambda$.  A cleaned map was restored
with a beam of 17\arcsec{} $\times$ 16\arcsec and background
subtraction was done, using the scaled 2cm VLA map, as described for
the Nobeyama 45 m telescope below.

\subsection{Nobeyama 45m}
The Nobeyama 45 m single dish observed the GC at two epochs using SIS
receiver at 40 and 100 GHz bands, and the Nobeyama bolometer array
(NOBA -- a seven element bolometer array; Kuno et al. 1993) at
150GHz. Except for the 150 GHz observations, the sources were observed
with cross scans and beam switching was used, which gives a beam throw
of 6\arcmin{}.  In addition to Sgr A* we observed Uranus (as primary
flux density calibrator), NRAO~530, and OH~5.89-0.39.

For May 24-28, 1996 we calibrated our flux densities against Uranus
(11.0Jy, 130K) for the 107 GHz and NGC~7027 (5.4 Jy) for the 43 GHz
observations.  The zenith optical depth was 0.1-0.2. Beam sizes were
15.2\arcsec{}$\times$16.2\arcsec{} at 107 GHz, and
36.5\arcsec{}$\times$ 40.7\arcsec{} 43 GHz. The relative uncertainty
of each observation is 0.2 Jy and during this observing run we did not
see any change in flux density of Sgr A* above the 10\% level. We
estimated the thermal contribution to the flux density within our beam
from the 15 GHz VLA data used by Zylka \& Mezger (1988). The VLA data
was scaled to the respective frequencies, using a spectral index of
$-0.1$ for free-free emission, and convolved with a beam corresponding
to the single dish observations.

For 25-30 Oct 1997 day-to-day variations of Sgr~A* were less than 15\%
of the total flux density at 95 GHz and less than 20\% at 150 GHz, hence we
only present flux densities averaged over several days (27, 29, \& 30 Oct for
95 GHz, and 25 \& 29 Oct for 150 GHz data). Beam sizes for NOBA at 150
GHz are 12\arcsec{} with separations of 16\arcsec{} and bridge readout 
technique was incorporated (Kuno et al. 1993). The flux densities of the
observed sources are given in Table \ref{Nobeyama}.

\subsection{IRAM 30 m}
We observed Sgr A* during the campaign with the IRAM 30 m single dish
antenna simultaneously with 3 receivers at 106.3, 152.3, and 235.6 GHz
and with beamsizes of 23\arcsec{}, 15\farcs7, and 10\farcs2
respectively. We had relatively bad weather on 26 October 1996, usable
conditions on 25, and very good weather on 27 October 1996 with an
optical depth of $\tau=0.2$ at 235.6 GHz. Chopping was used (beam
throw of 60\arcsec{}). To account for the effects of anomalous refraction,
each scan was shifted to the center position of fitted Gaussians with
shifts up to 3\arcsec{}. The contribution from thermal free-free
emission, as derived from the 15 GHz VLA map used in Zylka \& Mezger
(1988), was fitted to the wings of each scan and subtracted. A
statistical error of the measurement was derived from the standard
deviation of the combined scans. Flux densities and errors for Sgr~A*
are given in Table
\ref{fluxes}.

\section{Results}
Despite several lost datasets we were able to obtain a spectrum of Sgr
A* which covers the range from 1.36 to 232 GHz. The individual flux densities
are summarized in Table \ref{fluxes}.

\subsection{cm-Spectrum}
The spectrum at lower frequencies between 1.36 and 43 GHz was
successfully measured on two days by the VLA and is described by two
power laws with spectral indices $\alpha=0.17$
($S_\nu\propto\nu^\alpha$) below and $\alpha=0.30$ above 10 GHz, while
there is a marked break between 8.5 and 15 GHz with $\alpha=0.77$
(Fig.~1). We did not find any variability at and below 8.5 GHz between
the two days (at the 1\% level for 8.5GHz). We find an increase by
$\sim$15\% in flux at 15 GHz between 26 \& 27 Oct.~1996, however, in
addition to the statistical errors given in Tab.~2, we may have a
systematic error of up to $\sim$10\% at frequencies 15 GHz and higher
in the observations on 26 Oct 1996 due to possible improper correction
of different air masses. Thus we cannot claim any significant
intra-day variability of Sgr A* from our data at cm wavelengths.

\subsection{mm-Spectrum}
The errors of the mm-telescopes are much larger than those of the VLA
and comparison of various data-sets indicate that the variability at
$\lambda$3 \& 2mm is not larger than 20\% over a period of three days
during the campaign at all frequencies below 150 GHz (see also Gwinn
et al.~1991). We therefore first combined all available data for each
telescope to obtain an average flux density measurement of Sgr~A* over
the campaign period. Due to the different sampling of the datasets,
the mean of the average flux density will differ by $\pm$1 day between
the telescopes and in general is skewed towards the end of the
campaign period. In light of the limits of variability for Sgr A*, we
find that such an uncertainty is tolerable for the compilation of a
simultaneous, broad-band spectrum.

Since the individual flux densites from the three mm-telescopes agree
within the errors with each other we then combined the flux density
measurements from the different telescopes at $\lambda$3 \& 2mm and
compared it with the time-averaged VLA flux densities (Table 2,
Fig. 1).

First of all we notice that the $\lambda$3mm flux density is only
slightly above the extrapolation from the VLA observations which is
reassuring concerning the possible systematic uncertainties in the
thermal background subtraction. Hence, the fact that the $\lambda$2mm
flux density --- measured by the same telescopes --- lies
substantially (0.8Jy, $\sim4\sigma$) above the extrapolation from the
VLA cm data becomes even more significant. If we only consider
$\lambda$7, 3, \& 2mm we find that the spectral index of Sgr A*
increases to $\alpha=0.52$ in the mm-range, while the
$\lambda$3-to-2mm spectral index even becomes $\alpha=0.76$, hence
{\em we conclude that there is a significant mm-excess in the spectrum
of Sgr A~*}. The $\lambda$1.3mm measurement is consistent with such an
excess but, because of the large errors, has no significance for this
discussion.

\section{Summary \& Discussion}
A major problem in discussing the spectrum of Sgr A* and the reality
of the mm-excess has been the non-simultaneity of the measurements and
the comparison of array (low frequencies) and single dish (high
frequencies) flux densities. Our simultaneous observations now show a
smooth transition from the VLA to the mm-telescopes and an upturn in
the spectrum of Sgr~A* between $\lambda$3mm and $\lambda$2mm in the
single dish observations alone. The results of all telescopes were
consistent with each other and we conclude that the observed mm-excess
is not due to variability or technical artifacts.

A concern that needs to be addressed in more detailed when studying
the spectrum of Sgr A~* is, however, confusion by other sources. The
diffuse free-free emission in the GC obviously is a major source of
confusion for single dish observations of Sgr A* and subtraction of
this component was taken care of as discussed in Section 2. Optically
thick thermal emission, e.g.~from cold dust, on a scale of a few
arcseconds and beyond seems to be negligible in the mm-regime (Zylka
et al. 1995) and comparison of single-dish and high-resolution
interferometer flux density measurements (with OVRO) do not indicate
the presence of such components at smaller scales as discussed by
Serabyn et al.~(1992\footnote{The claimed possible counterpart
to IRS~7 at 222 GHz turned out to be an artifact of the image
processing (Serabyn et al.~1997).}, 1997).

Finally, a major contribution of a yet unknown non-variable point
source in the very vicinity of Sgr A* can also be ruled out.  Based on
high resolution VLA images with a resolution of 0\farcs1 at
$\lambda$13 and $\lambda$7 mm, the peak flux density for such a
hypothetical source is below 5 mJy at $\lambda$7mm.  The brightest
regions near Sgr A* are IRS 13 (3\arcsec{} SW to Sgr A*) which
contributes 100 mJy (integrated within a 3\arcsec$\times$3\arcsec{}
area) and IRS 2 (1-2\arcsec{} south to IRS 13) which contributes 85
mJy (within a 2\arcsec$\times$2\arcsec area)---the spectra of these
sources are consistent with optically thin free-free emission. From
the upper limits at lower frequencies it follows that any source which
might be confused with Sgr~A* would contribute negligibly to the total
flux density of Sgr A* at mm-wavelengths. For a source with optically
thick, thermal emission, for example, the contribution has to be less
than 50 mJy at $\lambda$2mm in order to be below the upper limits at
$\lambda$7mm. We therefore conclude that the mm-excess is indeed an
intrinsic feature of Sgr A*.

The sub-mm bump causing this excess is, in fact, very well explained
by assuming the presence of a compact, self-absorbed synchrotron
component in Sgr~A*. As outlined in Falcke (1996b; see also Beckert \&
Duschl 1997) this sub-mm component can be described in its most simple
minded form by four parameters: magnetic field $B$, electron density
$n$, electron Lorentz factor $\gamma_{\rm e}$, and radius $R$. These
input parameters correspond to three measured quantities: synchrotron
self-absorption (i.e. upturn) frequency $\nu_{\rm ssa}$, peak flux
density $S_{\nu_{\max}}$, and peak frequency $\nu_{\rm max}$. With an
equipartition parameter $k$ which is assumed to be of order unity (but
does not enter strongly) and a distance to the GC of 8.5 kpc, we find
that the radius of this emitting region is then given by

$$
R\sim1.5\cdot10^{12}{\rm cm}\quad
k^{-1/17} \left({S_{\nu_{\max}}\over 3.5 {\rm Jy}}\right)^{8/17}
$$
\begin{equation}
\left({\nu_{\rm max}\over 10^{12} {\rm Hz}}\right)^{-16/51}
\left({\nu_{\rm ssa}\over 100{\rm GHz}}\right)^{-35/51}.
\end{equation}

This size is consistent with the upper limits ($\sim$1~AU) from VLBI
(Rogers et al.~1994) and lower limits ($\sim10^{12}$cm) from
scintillation experiments (Gwinn et al. 1991). In comparison we also
note that the Schwarzschild radius ($R_{\rm S}$) of the putative
$2.6\cdot10^6 M_{\sun}$ black hole in the GC is already $R_{\rm
S}=0.77\cdot10^{12}$ cm and thus the compact sub-mm component should
correspond to a region in the very vicinity of the black hole. Most
interesting is the possibility that this region is directly affected
by general relativistic effects, and could for example be
gravitationally amplified if the radiation is intriniscially
anisotropic (e.g. similar to Cunningham 1975).

Finally, a compact component with the parameters as in Eq.~1 would be
very interesting for mm-VLBI, since, if this is not just a spot near
the black hole horizon, the black hole horizon could be imaged against
the background of this sub-mm emission. As Bardeen (1973, his Fig.~6)
has shown, a black hole in such a configuration would appear as a dark
disk with a diameter of 4.5 $R_{\rm S}$, (i.e. $3.45\cdot10^{12}$ cm,
or 27$\mu$as). Since the extrapolated scattering size of Sgr~A* at 215
GHz is also $\sim$27$\mu$as (e.g.~Yusef-Zadeh et al.~1994)---similar
to the expected resolution of future mm-VLBI experiments (e.g.~Wright
\& Bower 1997), direct imaging of the putative black hole in the GC
might be in reach within the next decade.

\acknowledgements 
Observations of Sgr A* were performed by MG \& J.-H.Z (VLA), RZ
(IRAM), PT \& HF (BIMA), HM (Nobeyama).  HM is grateful to Hideaki
Kashihara and Akihiro Sakamoto for their help during observations.
This research was supported in part by NASA under grants NAGW-3268,
NAG8-1027, by NSF under grant AST 96-13716, and by the DFG under grant
Fa 358/1-1\&2. The National Radio Astronomy Observatory is a facility
of the National Science Foundation, operated under a cooperative
agreement by Associated Universities, Inc. Nobeyama Radio Observatory
(NRO) is a branch of the National Astronomical Observatory, an
inter-university research institute operated by the Ministry of
Education, Science, Sports and Culture.


\clearpage
\renewcommand{\baselinestretch}{1}\small

\begin{deluxetable}{lrrrr}
\tablecaption{Nobeyama Flux Densities}
\tablehead{
\colhead{Source}  &
\colhead{107 GHz} &
\colhead{ 43 GHz} &
\colhead{150 GHz} &
\colhead{ 95 GHz} 
}
\tablecolumns{3}
\startdata
NRAO~530 	&   13.2  $\pm$ 0.4   Jy  &16.0  $\pm$ 0.5 Jy& 12.0  $\pm$ 1.2   Jy  &    12.6  $\pm$ 0.7 Jy\\
OH~5.89--0.39	&    8.5  $\pm$ 0.3   Jy  & 9.0  $\pm$ 0.3 Jy&  8.8 $\pm$ 0.9    Jy  &     8.0  $\pm$ 0.5 Jy\\
Sgr~A*+thermal 	&    5.5  $\pm$ 0.2   Jy  &11.0  $\pm$ 0.3 Jy&  4.0  $\pm$ 0.4   Jy  &     5.1  $\pm$ 0.3 Jy\\
Sgr~A*  	&    2.7  $\pm$ 0.4   Jy  & 1.9  $\pm$ 0.6 Jy&  3.1  $\pm$ 0.4   Jy  &     2.0  $\pm$ 0.4 Jy
\tablecomments{Flux densities and 1 $\sigma$ errors for
Galactic Center sources obtained with the Nobeyama 45m antenna. 107
and 43 GHz flux densites are from 24-28 May 1996, and 150 and 95 GHz
flux densities are from 25-30 Oct.~1996. 
}
\label{Nobeyama}
\enddata
\end{deluxetable}

\begin{deluxetable}{llllllll}
\tablecaption{Measured Flux Densities for Sgr A*}
\tablehead{ 
\colhead{}&
\colhead{}&
\multicolumn{2}{c}{\hrulefill{}26 Oct 1996\hrulefill{}}&
\multicolumn{2}{c}{\hrulefill{}27 Oct 1996\hrulefill{}}&
\multicolumn{2}{c}{\hrulefill{}All days\hrulefill{}}\\
\colhead{telescope}&
\colhead{$\nu$}&
\colhead{$S_{\nu}$}&
\colhead{err.}&
\colhead{$S_{\nu}$}&
\colhead{err.}&
\colhead{$S_{\nu}$}&
\colhead{err.}\\
\colhead{}&
\colhead{[GHz]}&
\multicolumn{2}{c}{[Jy]}&
\multicolumn{2}{c}{[Jy]}&
\multicolumn{2}{c}{[Jy]}
}
\tablecolumns{8}
\startdata
IRAM 		& 235.6	&     	&    	&      	&    	& 2.8	& $\pm$0.9	\\
IRAM		& 152.3	&     	&     	&      	&     	& 2.9	& $\pm$0.25	\\
IRAM+Nobeyama	& 151.	& 	&	&	& 	& 2.9	& $\pm$0.21	\\
Nobeyama	& 150.	&    	&    	&     	&    	& 3.1	& $\pm$0.4	\\
IRAM		& 106.3	&     	&     	&     	&     	& 2.4	& $\pm$0.35	\\
IRAM$^\dagger$+Nobeyama+BIMA	
  		&  95.0	& 	&	&	&    	& 2.1	& $\pm$0.23 	\\
Nobeyama	          &  95. 	&    	&    	&     	&    	& 2.0	& $\pm$0.4	\\
BIMA		&  93. 	&     	&    	&     	&    	& 1.9	& $\pm$0.5	\\
VLA		&  43.3	& 1.4	& $\pm$0.2	& 1.50 	& $\pm$0.05	& 1.3	& $\pm$0.14	\\
VLA		&  22.5	& 1.0 	& $\pm$0.1	& 1.20 	& $\pm$0.07	& 1.1	& $\pm$0.05	\\
VLA		&  14.9	& 0.95	& $\pm$0.06	& 1.10 	& $\pm$0.03	& 1.03	& $\pm$0.03	\\
VLA		&   8.45& 0.72	& $\pm$0.01	& 0.71	& $\pm$0.01	& 0.72	& $\pm$0.01	\\
VLA		&   4.85&  	& 	& 0.64	& $\pm$0.01	& 0.64	& $\pm$0.01	\\
VLA		&   1.64&  	& 	& 0.55	& $\pm$0.03	& 0.55	& $\pm$0.03	\\
VLA		&   1.44& 0.54	& $\pm$0.04	& 0.52	& $\pm$0.02	& 0.53	& $\pm$0.02	\\
VLA		&   1.36& 0.53	& $\pm$0.04	& 0.52	& $\pm$0.02	& 0.53	& $\pm$0.02	
\tablecomments{Description
of columns: (1) -- telescope or combination of telescope involved to
derive the flux density, (2) -- frequency (average if multiple
telescopes), (3) -- VLA flux density on 26 Oct. 1996, (4) -- one $\sigma$
error, (5) -- VLA flux density on 27 Oct. 1996, (6) -- one $\sigma$ error, (7)
-- flux density averaged over the available data from all three days
(linear average of columns 3 \& 5 for VLA data else weighted by
error), (8) -- combined one
$\sigma$ error. Additional note: $^\dagger$ calculated from 106.3 GHz
assuming a spectral index of $\alpha=0.76$.}
\label{fluxes}
\enddata
\end{deluxetable}

\onecolumn
\begin{figure}
\plotone{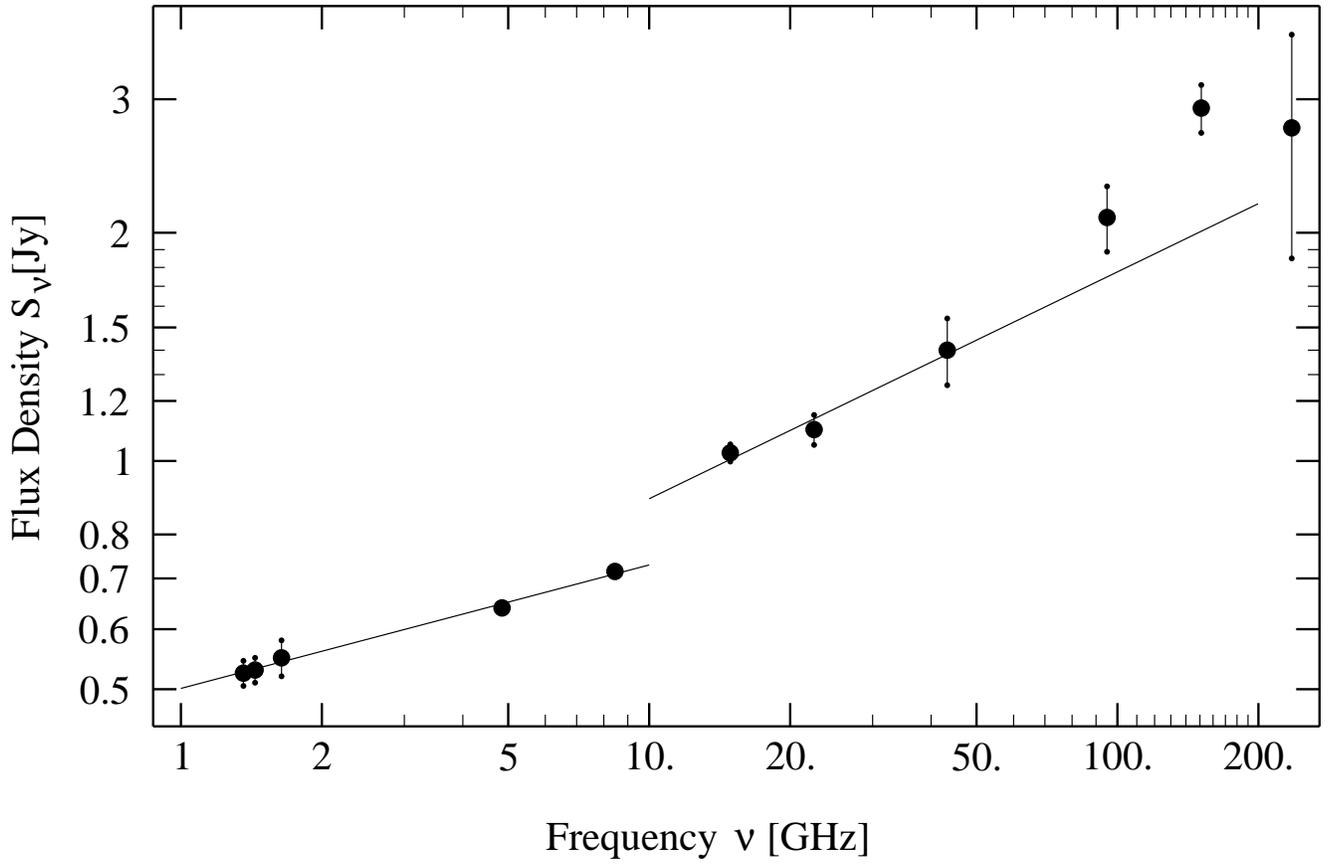}
\caption{\label{spectrum2}Spectrum of Sgr~A* plotted as the
logarithm of the flux density vs. the logarithm of the
frequency. Shown are the data averaged over the campaign period; flux
densities at neighboring frequencies were also combined from the
different mm-telescopes. Solid lines represent power law fits to the
low- and high-frequency VLA data.}
\end{figure}
\end{document}